\RequirePackage{lineno}
\documentclass[aps,prd,twocolumn,superscriptaddress,showpacs,amsmath,amssymb]{revtex4}

\usepackage{comment}
\usepackage{multirow}
\usepackage{dcolumn}
\usepackage{bm}
\usepackage{setspace}
\usepackage{graphicx}
\includecomment{printrobustnotes}
\includecomment{printallnotes}
\usepackage{xspace}

\newenvironment{cfigure1c}[1][tbp]{\begin{figure*}[#1]\centering}{\end{figure*}}

\newcommand{\note}[1]{\xspace}
\newcommand{\robustnote}[1]{\xspace}
\begin{printrobustnotes}
\renewcommand{\robustnote}[1]{\textbf{\textit{#1}}\xspace}
\end{printrobustnotes}
\begin{printallnotes}
\renewcommand{\note}[1]{\textbf{\textit{#1}}\xspace}
\renewcommand{\robustnote}[1]{\textbf{\textit{#1}}\xspace}
\end{printallnotes}
\newcommand{\ppbar}  {\ensuremath{p\bar{p}}\xspace}
\newcommand{\ttbar}  {\ensuremath{t\bar{t}}\xspace}

\newcommand{\et}     {\ensuremath{E_{T}}\xspace}

\newcommand{\pt}     {\ensuremath{p_{T}}\xspace}

\newcommand{\met}    {\ensuremath{\text{\raisebox{.3ex}{$\not$}}\et}\xspace}

\newcommand{\ma}    {\ensuremath{m_{t}^{\text{had,reco}}}\xspace}
\newcommand{\mb}    {\ensuremath{m_{t}^{\text{had,reco(2)}}}\xspace}

\newcommand{\mtop}    {\ensuremath{\mathrm{M}_{\text{top}}}\xspace}

\newcommand{\mjj}    {\ensuremath{m_{jj}}\xspace}

\newcommand{\genunit}[2]{\ensuremath{#1~\mathrm{#2}}\xspace}

\newcommand{\gev}[1]    {\genunit{#1}{GeV}}

\newcommand{\gevc}[1]   {\ensuremath{#1~\mathrm{GeV}/c}}
\newcommand{\gevcc}[1]  {\ensuremath{#1~\mathrm{GeV}/c^{2}}}

\newcommand{\invpb}[1]  {\ensuremath{#1~\mathrm{pb}^{-1}}}

\newcommand{\measErr}[2]{\ensuremath{#1 \pm #2}\xspace}

\newcommand{\measStatSyst}[3]{\ensuremath{#1 \pm #2~(\textrm{stat}) \pm #3~(\textrm{syst})}\xspace}

\begin{document}

\setpagewiselinenumbers 
\modulolinenumbers[1] 


\title{{\bf Top-quark mass measurement using events with missing transverse energy and jets at CDF}}

\affiliation{Institute of Physics, Academia Sinica, Taipei, Taiwan 11529, Republic of China} 
\affiliation{Argonne National Laboratory, Argonne, Illinois 60439, USA} 
\affiliation{University of Athens, 157 71 Athens, Greece} 
\affiliation{Institut de Fisica d'Altes Energies, ICREA, Universitat Autonoma de Barcelona, E-08193, Bellaterra (Barcelona), Spain} 
\affiliation{Baylor University, Waco, Texas 76798, USA} 
\affiliation{Istituto Nazionale di Fisica Nucleare Bologna, $^{aa}$University of Bologna, I-40127 Bologna, Italy} 
\affiliation{University of California, Davis, Davis, California 95616, USA} 
\affiliation{University of California, Los Angeles, Los Angeles, California 90024, USA} 
\affiliation{Instituto de Fisica de Cantabria, CSIC-University of Cantabria, 39005 Santander, Spain} 
\affiliation{Carnegie Mellon University, Pittsburgh, Pennsylvania 15213, USA} 
\affiliation{Enrico Fermi Institute, University of Chicago, Chicago, Illinois 60637, USA}
\affiliation{Comenius University, 842 48 Bratislava, Slovakia; Institute of Experimental Physics, 040 01 Kosice, Slovakia} 
\affiliation{Joint Institute for Nuclear Research, RU-141980 Dubna, Russia} 
\affiliation{Duke University, Durham, North Carolina 27708, USA} 
\affiliation{Fermi National Accelerator Laboratory, Batavia, Illinois 60510, USA} 
\affiliation{University of Florida, Gainesville, Florida 32611, USA} 
\affiliation{Laboratori Nazionali di Frascati, Istituto Nazionale di Fisica Nucleare, I-00044 Frascati, Italy} 
\affiliation{University of Geneva, CH-1211 Geneva 4, Switzerland} 
\affiliation{Glasgow University, Glasgow G12 8QQ, United Kingdom} 
\affiliation{Harvard University, Cambridge, Massachusetts 02138, USA} 
\affiliation{Division of High Energy Physics, Department of Physics, University of Helsinki and Helsinki Institute of Physics, FIN-00014, Helsinki, Finland} 
\affiliation{University of Illinois, Urbana, Illinois 61801, USA} 
\affiliation{The Johns Hopkins University, Baltimore, Maryland 21218, USA} 
\affiliation{Institut f\"{u}r Experimentelle Kernphysik, Karlsruhe Institute of Technology, D-76131 Karlsruhe, Germany} 
\affiliation{Center for High Energy Physics: Kyungpook National University, Daegu 702-701, Korea; Seoul National University, Seoul 151-742, Korea; Sungkyunkwan University, Suwon 440-746, Korea; Korea Institute of Science and Technology Information, Daejeon 305-806, Korea; Chonnam National University, Gwangju 500-757, Korea; Chonbuk National University, Jeonju 561-756, Korea} 
\affiliation{Ernest Orlando Lawrence Berkeley National Laboratory, Berkeley, California 94720, USA} 
\affiliation{University of Liverpool, Liverpool L69 7ZE, United Kingdom} 
\affiliation{University College London, London WC1E 6BT, United Kingdom} 
\affiliation{Centro de Investigaciones Energeticas Medioambientales y Tecnologicas, E-28040 Madrid, Spain} 
\affiliation{Massachusetts Institute of Technology, Cambridge, Massachusetts 02139, USA} 
\affiliation{Institute of Particle Physics: McGill University, Montr\'{e}al, Qu\'{e}bec, Canada H3A~2T8; Simon Fraser University, Burnaby, British Columbia, Canada V5A~1S6; University of Toronto, Toronto, Ontario, Canada M5S~1A7; and TRIUMF, Vancouver, British Columbia, Canada V6T~2A3} 
\affiliation{University of Michigan, Ann Arbor, Michigan 48109, USA} 
\affiliation{Michigan State University, East Lansing, Michigan 48824, USA}
\affiliation{Institution for Theoretical and Experimental Physics, ITEP, Moscow 117259, Russia}
\affiliation{University of New Mexico, Albuquerque, New Mexico 87131, USA} 
\affiliation{Northwestern University, Evanston, Illinois 60208, USA} 
\affiliation{The Ohio State University, Columbus, Ohio 43210, USA} 
\affiliation{Okayama University, Okayama 700-8530, Japan} 
\affiliation{Osaka City University, Osaka 588, Japan} 
\affiliation{University of Oxford, Oxford OX1 3RH, United Kingdom} 
\affiliation{Istituto Nazionale di Fisica Nucleare, Sezione di Padova-Trento, $^{bb}$University of Padova, I-35131 Padova, Italy} 
\affiliation{LPNHE, Universite Pierre et Marie Curie/IN2P3-CNRS, UMR7585, Paris, F-75252 France} 
\affiliation{University of Pennsylvania, Philadelphia, Pennsylvania 19104, USA}
\affiliation{Istituto Nazionale di Fisica Nucleare Pisa, $^{cc}$University of Pisa, $^{dd}$University of Siena and $^{ee}$Scuola Normale Superiore, I-56127 Pisa, Italy} 
\affiliation{University of Pittsburgh, Pittsburgh, Pennsylvania 15260, USA} 
\affiliation{Purdue University, West Lafayette, Indiana 47907, USA} 
\affiliation{University of Rochester, Rochester, New York 14627, USA} 
\affiliation{The Rockefeller University, New York, New York 10065, USA} 
\affiliation{Istituto Nazionale di Fisica Nucleare, Sezione di Roma 1, $^{ff}$Sapienza Universit\`{a} di Roma, I-00185 Roma, Italy} 

\affiliation{Rutgers University, Piscataway, New Jersey 08855, USA} 
\affiliation{Texas A\&M University, College Station, Texas 77843, USA} 
\affiliation{Istituto Nazionale di Fisica Nucleare Trieste/Udine, I-34100 Trieste, $^{gg}$University of Udine, I-33100 Udine, Italy} 
\affiliation{University of Tsukuba, Tsukuba, Ibaraki 305, Japan} 
\affiliation{Tufts University, Medford, Massachusetts 02155, USA} 
\affiliation{University of Virginia, Charlottesville, Virginia 22906, USA}
\affiliation{Waseda University, Tokyo 169, Japan} 
\affiliation{Wayne State University, Detroit, Michigan 48201, USA} 
\affiliation{University of Wisconsin, Madison, Wisconsin 53706, USA} 
\affiliation{Yale University, New Haven, Connecticut 06520, USA} 
\author{T.~Aaltonen}
\affiliation{Division of High Energy Physics, Department of Physics, University of Helsinki and Helsinki Institute of Physics, FIN-00014, Helsinki, Finland}
\author{B.~\'{A}lvarez~Gonz\'{a}lez$^w$}
\affiliation{Instituto de Fisica de Cantabria, CSIC-University of Cantabria, 39005 Santander, Spain}
\author{S.~Amerio}
\affiliation{Istituto Nazionale di Fisica Nucleare, Sezione di Padova-Trento, $^{bb}$University of Padova, I-35131 Padova, Italy} 

\author{D.~Amidei}
\affiliation{University of Michigan, Ann Arbor, Michigan 48109, USA}
\author{A.~Anastassov}
\affiliation{Northwestern University, Evanston, Illinois 60208, USA}
\author{A.~Annovi}
\affiliation{Laboratori Nazionali di Frascati, Istituto Nazionale di Fisica Nucleare, I-00044 Frascati, Italy}
\author{J.~Antos}
\affiliation{Comenius University, 842 48 Bratislava, Slovakia; Institute of Experimental Physics, 040 01 Kosice, Slovakia}
\author{G.~Apollinari}
\affiliation{Fermi National Accelerator Laboratory, Batavia, Illinois 60510, USA}
\author{J.A.~Appel}
\affiliation{Fermi National Accelerator Laboratory, Batavia, Illinois 60510, USA}
\author{A.~Apresyan}
\affiliation{Purdue University, West Lafayette, Indiana 47907, USA}
% ...
\collaboration{CDF Collaboration\footnote{With visitors from $^a$Istituto Nazionale di Fisica Nucleare, Sezione di Cagliari, 09042 Monserrato (Cagliari), Italy,
$^b$University of CA Irvine, Irvine, CA  92697, USA,
$^c$University of CA Santa Barbara, Santa Barbara, CA 93106, USA,
$^d$University of CA Santa Cruz, Santa Cruz, CA  95064, USA,
$^e$CERN,CH-1211 Geneva, Switzerland,
$^f$Cornell University, Ithaca, NY  14853, USA, 
$^g$University of Cyprus, Nicosia CY-1678, Cyprus, 
$^h$Office of Science, U.S. Department of Energy, Washington, DC 20585, USA,
$^i$University College Dublin, Dublin 4, Ireland,
$^j$University of Fukui, Fukui City, Fukui Prefecture, Japan 910-0017,
$^k$Universidad Iberoamericana, Mexico D.F., Mexico,
$^l$Iowa State University, Ames, IA  50011, USA,
$^m$University of Iowa, Iowa City, IA  52242, USA,
$^n$Kinki University, Higashi-Osaka City, Japan 577-8502,
$^o$Kansas State University, Manhattan, KS 66506, USA,
$^p$University of Manchester, Manchester M13 9PL, United Kingdom,
$^q$Queen Mary, University of London, London, E1 4NS, United Kingdom,
$^r$University of Melbourne, Victoria 3010, Australia,
$^s$Muons, Inc., Batavia, IL 60510, USA,
$^t$Nagasaki Institute of Applied Science, Nagasaki, Japan, 
$^u$National Research Nuclear University, Moscow, Russia,
$^v$University of Notre Dame, Notre Dame, IN 46556, USA,
$^w$Universidad de Oviedo, E-33007 Oviedo, Spain, 
$^x$Texas Tech University, Lubbock, TX  79609, USA,
$^y$Universidad Tecnica Federico Santa Maria, 110v Valparaiso, Chile,
$^z$Yarmouk University, Irbid 211-63, Jordan,
$^{hh}$On leave from J.~Stefan Institute, Ljubljana, Slovenia, 
}}
\noaffiliation


\begin{abstract}
We present a measurement of the top-quark mass using a sample of $t\bar{t}$ events in 5.7 fb$^{-1}$ of integrated luminosity from $p\bar{p}$ collisions at the Fermilab Tevatron with $\sqrt{s} =$ 1.96 TeV and collected by the CDF II Detector. We select events having no identified charged leptons, large missing transverse energy, and four, five, or six jets with at least one jet tagged as coming from a $b$ quark. This analysis considers events from the semileptonic \ttbar decay channel, including events that contain tau leptons. The measurement is based on a multidimensional template method. We fit the data to signal templates of varying top-quark masses and background templates, and measure a top-quark mass of $\mtop = \gevcc{\measStatSyst{172.3}{2.4}{1.0}}$.
\end{abstract}
\maketitle
The top quark~($t$) is the heaviest known elementary particle, with a mass approximately 40 times heavier than the mass of the bottom quark~($b$), its partner in the weak isospin doublet. The top-quark mass can be used as a consistency check for standard model~(SM) parameters. For example, the top quark contributes significantly to electroweak radiative corrections relating its mass~(\mtop) and that of the $W$ boson to the mass of the predicted Higgs boson within the SM and in new physics models~\cite{hmass,hmass1}. Therefore, combined precision measurements of the $W$ mass and \mtop provide an important constraint on the Higgs boson mass. At the Tevatron, the top quark is predominantly produced in \ttbar pairs. As in the SM the top quark decays almost exclusively to a $W$ boson and a $b$ quark, the expected signature of a \ttbar production event is $t\bar{t} \rightarrow W^{+}bW^{-}\bar{b}$, assuming unitarity of the three-generation CKM matrix~\cite{ckm}. Since the $W$ boson subsequently decays either to a quark-antiquark pair or to a lepton-neutrino pair, the final state of $\ttbar$ production can be classified by the number of charged leptons produced. In this letter, we focus on events with large missing transverse energy~(\met)~\cite{DefMET} as expected for undetected energetic neutrinos, accompanied by jets. We explicitly veto events with identified high $p_T$ electrons or muons~(lepton+jets events) as well as multijet events where both $W$ bosons decay hadronically~(all-hadronic events). This ensures that our result is statistically independent from other CDF top-quark mass measurements~\cite{cdflj2,cdflj,cdfdil,cdfall} and allows for a future combination with them. A previous measurement of \mtop in this final state used an integrated luminosity of \invpb{311}~\cite{earlyMETJ} and yielded $\mtop = \gevcc{\measStatSyst{172.3}{15.3}{14.4}}$. Although no identified leptons are explicitly required, our measurement is sensitive to all $W$ leptonic decays. This includes decays to $\tau$ leptons, which constitute approximately 40\% of the signal sample in our final selection. The frequency of observed $\tau$  lepton final states is predicted to be enhanced by new physics models such as a charged Higgs decay~\cite{chargedHiggs}, therefore, a significantly different measurement of \mtop in this decay channel could indicate contributions from non-SM~\cite{bSM} physics processes.

We use data corresponding to an integrated luminosity of 5.7 fb$^{-1}$ of \ppbar collisions at the Fermilab Tevatron, and collected by the CDF II Detector~\cite{cdfii}. The sample of events used in this measurement is a subset of events that initially passed a trigger requirement, which accepted events with at least four calorimetric clusters~\cite{calclu} of $E_{T} > \gev{15}$ and a scalar sum $E_T$ of these clusters greater than \gev{175}~\cite{METJcross}. After the trigger selection, event observables of physical interest are computed. Jets are reconstructed with the {\sc jetclu}~\cite{jetclu} algorithm using a cone radius of $\Delta R=\sqrt{{\Delta\eta}^2+{\Delta\phi}^2}=0.4$~\cite{cordSystem}. Jets are corrected~\cite{jetcorrect} for nonuniformities of the calorimeter response as a function of $\eta$, energy contributed by multiple \ppbar interactions in the event, and calorimeter nonlinear response. To determine if a jet originated from a $b$ quark, a secondary vertex algorithm~\cite{secvtx} is applied. This algorithm identifies jets that are likely to come from $b$ quark hadronization through the presence of a displaced vertex within the jets~($b$-tag). We require at least one jet to be identified as a $b$ jet~($b$-tagged). We divide the sample of candidate events into two, separating events with one $b$-tagged jet~(1-tag) from events with two or more $b$-tagged jets~(2-tag). Events are required to have at least four and at most six jets with transverse energy $\et > \gev{15}$ and $|\eta| <2.0$. To avoid overlap with other CDF top-quark mass measurements, we reject events with reconstructed electrons or muons with $\pt > \gevc{20}$ and $|\eta| < 1.0$~(lepton + jets final state), and events with \met significance below $\gev{3}^{1/2}$~(all-hadronic final state), where the \met significance is defined as ${\met}^{sig} = \met/\sqrt{\sum_{jets} E_T}$. For further rejection of multijet backgrounds from QCD processes, we require $\Delta\phi_{min}(\met,jet) > 0.4$, where $\Delta\phi_{min}(\met,jet)$ is the smallest separation in the angle $\phi$ between jets and \met~\cite{METJcross}.

Background events with $b$-tags arise from QCD multijet and electroweak production of $W$ bosons associated with heavy flavor jets. In order to improve the signal-to-background ratio~(S/B) in this analysis, an artificial neural network is trained to identify the kinematic and topological charateristics of SM \ttbar events using eight input variables~\cite{METJcross}. We apply the neural network to all events passing the above selections, and make a cut on neural network score that retains 81\% of the \ttbar signal events while rejecting 91\% of background events. The selection criteria listed above define the `signal region'. We follow \cite{METJcross} and estimate the background rate using a data-driven method. The method uses events with exactly three jets and employs a per-jet parameterization of the $b$-tagging probability. Due to the presence of \ttbar events in samples with higher jet multiplicity, we extrapolate the $b$-tagging probability of the three-jet event sample to higher jet multiplicity events by iteratively removing the \ttbar content from the sample~\cite{METJcross}. We estimate the background for the 1-tag and 2-tag samples separately. A $b$-tagging correction factor~\cite{tagcorr} is applied to take into account the fact that most of the heavy flavor jets are produced in pairs. With this procedure we obtain the estimated number of background events in the signal region shown in Table~\ref{nevent}. We also show the estimated number of \ttbar signal events, assuming a $\ttbar$ production cross-section of 7.5~pb at \mtop = \gevcc{172.5}~\cite{xseccorr}, together with the number of observed events in the data.

Monte Carlo~(MC)-simulated \ttbar samples are generated by {\sc pythia}~\cite{pythia} with 76 different \mtop values ranging from 150 GeV/c$^2$ to 240 GeV/c$^2$. For each of them we reconstruct the events with different values of $\Delta_{JES}$, the difference between the jet energy scale~(JES) in the MC simulation and the data. A total of 27 different $\Delta_{JES}$ values, ranging from -3.0 $\sigma_c$ to +3.0 $\sigma_c$, where $\sigma_c$ is the uncertainty on the JES~\cite{jes}, are used to reduce the systematic effects due to the jet energy uncertainty, as described later.

\begin{table}
\begin{center}
\caption[Number of events]{Number of expected signal and background events and observed data events with integrated luminosity of 5.7 fb$^{-1}$ in the signal region. The \ttbar signal contribution is estimated with a cross-section of 7.5 pb. All selection requirements are applied, and events are separated into 1-tag and 2-tag categories.}
\begin{ruledtabular}
\begin{tabular}{ccc}
\label{nevent}
Events &             1-tag & 2-tag \\
\hline           
Expected \ttbar Signal     & 644.3 $\pm$ 118.7 & 262.9 $\pm$ 50.3 \\
Expected Background & 410.6 $\pm$ 31.7 & 43.8  $\pm$ 11.0 \\
Total Expectation   & 1054.9 $\pm$ 122.9& 306.7 $\pm$ 51.5\\
Observed Data       & 1147 & 285
\end{tabular}
\end{ruledtabular}
\end{center}
\end{table}

For each MC-simulated sample in this analysis, we reconstruct three variables using the leading four or five jets to form the templates. The first variable, \mjj, defined as the invariant mass of the two jets from the hadronically decaying $W$  boson, serves as an {\it in situ} constraint of the JES through the likelihood fit described later. We calculate \mjj from the two non-$b$-tagged jets whose invariant mass produces the closest value to the world average $W$ boson mass of 80.40 GeV/c$^2$~\cite{wmass}. We also extract the reconstructed top-quark mass~(\ma) from the invariant mass of the three jets whose momentum sum yields the largest $p_T$ since the invariant mass constructed this way has a large correlation with the hadronically decaying top-quark mass. To enhance the probability that the jets used to compute \ma come from the hadronically decaying top quark, we add two constraints to the calculation of \ma: first, \ma must contain the two jets that contribute to \mjj, and second, in 2-tag events one jet of \ma must be $b$-tagged. A third variable, \mb, is defined as the invariant mass of three jets, two of which are the pair which defines \mjj. The third jet of \mb is required to be the most energetic jet of those not forming \ma, and is required to be $b$-tagged in a 2-tag event. The variable \mb provides complementary information on the top quark mass, and is  particularly important in events where the three jets used to compute \ma were not the actual decay products of the hadronically-decaying top quark.

The template method used in the extraction of \mtop requires that a probability density function~(p.d.f.) be built for each template. For each MC signal and background sample, we estimate the p.d.f.s using the kernel density estimation~(KDE)\cite{kde1,kde2} that employs a non-parametric method to construct p.d.f.s. For each sample, we build a three dimensional p.d.f. from the reconstructed observables~(\mjj, \ma, and \mb), taking their correlations into account. To measure \mtop, we fit the signal and background p.d.f.s to the distributions of the observables in the data using an unbinned maximum likelihood fit~\cite{likelihood_fit} where we minimize the negative logarithm of the likelihood using {\sc minuit}~\cite{minuit}. The likelihood fits \mtop and $\Delta_{JES}$ simultaneously, and is built separately for each subsample, 1-tag and 2-tag events, in order to improve the usage of statistical information. References~\cite{kde2,like1} provide detailed information about this technique.

The mass fitting procedure is tested with pseudoexperiments for a set of MC \ttbar samples with 14 different \mtop values ranging from 159 GeV/c$^2$ to 185 GeV/c$^2$. For each pseudoexperiment we select the number of background events from a Poisson distribution with a mean equal to the expected total number of background events in the sample and the number of signal events from a Poisson distribution with a mean equal to the expected number of signal events normalized to a \ttbar production cross section of 7.5 pb~\cite{xseccorr}. The distributions of the average mass residual (the deviation from the input top-quark mass) and the width of the pull (the ratio of the residual to the uncertainty) for simulated experiments are corrected to be zero and unity, respectively. The correction is $M_{t}^{corr} = 1.24 \times M_t^{meas}- 40.6$ GeV/c$^2$, where $M_t^{meas}$ is the raw value from the likelihood fit and $M_t^{corr}$ is the corrected value of the measurement. The measured uncertainty is correspondingly increased by 20\% to correct the width of the pull distribution. 

We examine various sources of systematic uncertainties that could affect the measurement by comparing the results of pseudoexperiments in which we vary relevant parameters within their uncertainties. One of the dominant sources of systematic uncertainty is the residual JES~\cite{jes,cdflj}. We vary the JES components within their uncertainties in the generated signal MC events and interpret the shifts in the returned top-quark mass as uncertainties. The $b$ jet energy scale systematic uncertainty that arises from the modeling of $b$ fragmentation, $b$ hadron branching fractions, and calorimeter response captures the additional uncertainty not taken into account in the light quark jet energy scale~\cite{cdflj}. The uncertainty arising from the choice of MC generator is estimated by comparing results from MC simulated samples generated with {\sc pythia} and {\sc herwig}~\cite{herwig}. We estimate the systematic uncertainty due to imperfect modeling of initial-state gluon radiation~(ISR) and final-state gluon radiation~(FSR) by varying the amounts of ISR and FSR in simulated events~\cite{LJtmass}. We estimate the systematic uncertainty due to parton distribution functions~(PDF's) of the proton by varying the independent eigenvectors of the CTEQ6M~\cite{cteq6} PDF's, varying $\Lambda_{QCD}$~(228 MeV vs. 300 MeV), and comparing CTEQ5M~\cite{cteq5} with MRST72~\cite{mrst72} PDF's. To estimate the systematic uncertainty associated with uncertainties in the top-quark production mechanism, we vary the fraction of the top quarks produced by gluon-gluon annihilation from the default 6$\%$ to 20$\%$, corresponding to a one standard deviation upper bound on the gluon fusion fraction~\cite{gg}. We also evaluate the uncertainty due to background modeling effects by re-weighting the background shape up and down and comparing the resulting measurements. We apply an additional uncertainty to account for the effect of the trigger simulation in the signal MC samples, in a similar way to the background shape systematic uncertainty estimation. We also estimate an uncertainty due to the effect of  multiple hadron interactions, which takes into account the increasing instantaneous luminosity in this dataset. The color reconnection (CR) systematic uncertainty~\cite{cr2} is evaluated using MC samples generated with and without CR effects adopting different tunes~\cite{mt2} of {\sc pythia}. Table~\ref{systtable} summarizes the individual systematic uncertainties considered, giving a total systematic uncertainty of 1.0 GeV/c$^2$ for the measurement of \mtop.

\begin{table}
\begin{center}
\caption[Summary of systematics]{Systematic uncertainties on \mtop.}
\begin{ruledtabular}
\begin{tabular}{cc}
\label{systtable}
Systematic Sources & Uncertainty~(GeV/c$^2$) \\
\hline
Residual JES           & 0.5 \\
$b$-JES           &0.3 \\
MC Generator             & 0.7 \\
Radiation              &0.2 \\
PDFs                   &0.2 \\
gg fraction            &$<$ 0.1 \\
Background        &0.1 \\
Trigger Modeling        & 0.1 \\
Multiple Hadron Interaction   &0.2 \\
Color Reconnection     &0.2 \\
\hline                      
Total Systematic Uncertainty       &1.0\\
\end{tabular}
\end{ruledtabular}
\end{center}
\end{table}

By applying a likelihood fit to the data using the three observables described above and the corrections obtained from the simulated experiments, the top-quark mass is measured to be 
\begin{eqnarray}
\mtop & = & 172.3 \pm 2.4 (stat) \pm 1.0 (syst) \gevcc \nonumber \\
      & = & 172.3 \pm 2.6  \gevcc  \,.
\end{eqnarray}

Figure~\ref{fig:pdf1d} shows the distribution of the observables used for the \mtop measurement overlaid with their probability density functions from \ttbar signal events with \mtop = 172.5 GeV/c$^2$ and the estimated background.

\begin{cfigure1c}
\begin{tabular}{ccc}
\includegraphics[width=0.33\textwidth]{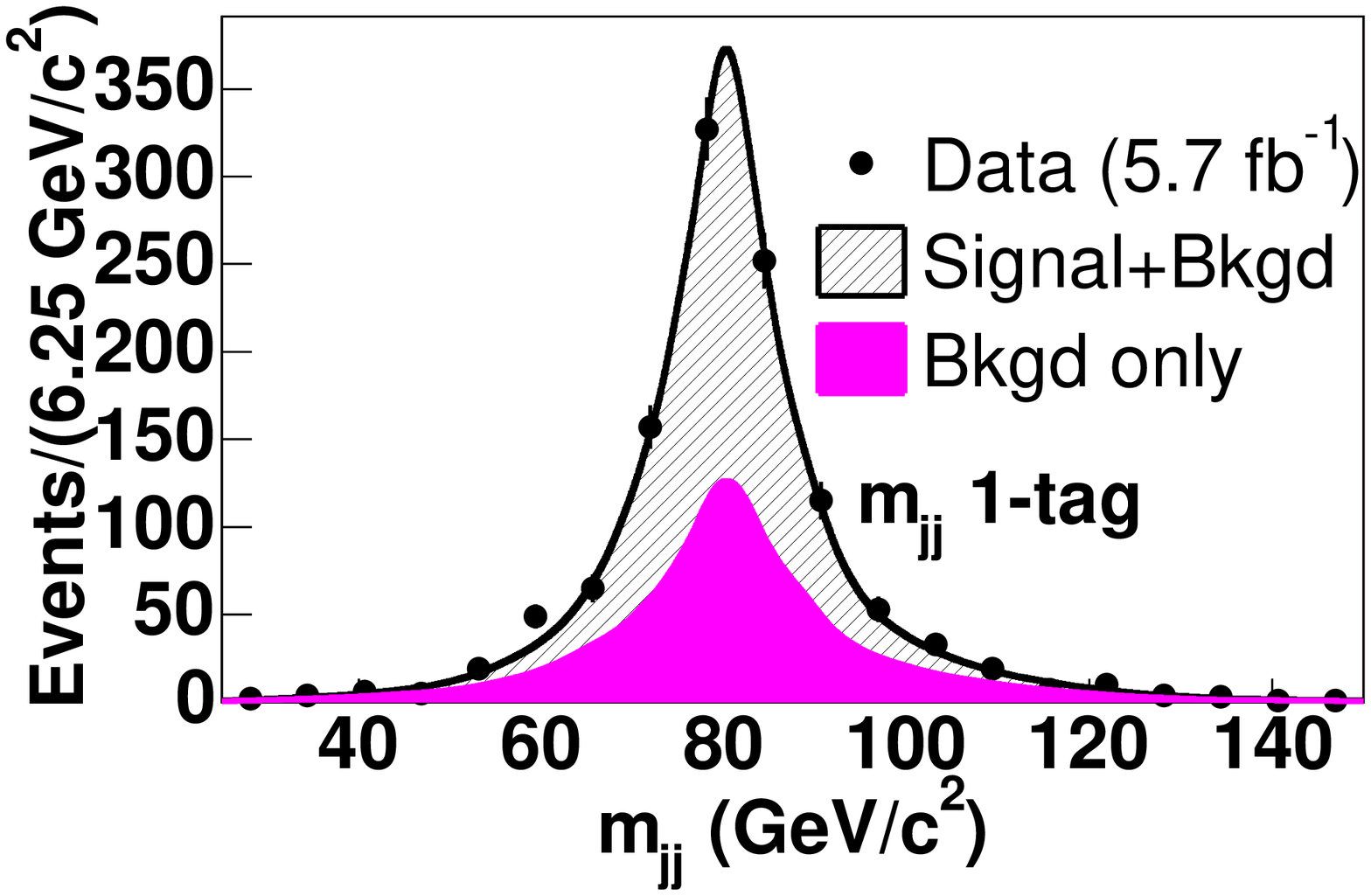} &
\includegraphics[width=0.33\textwidth]{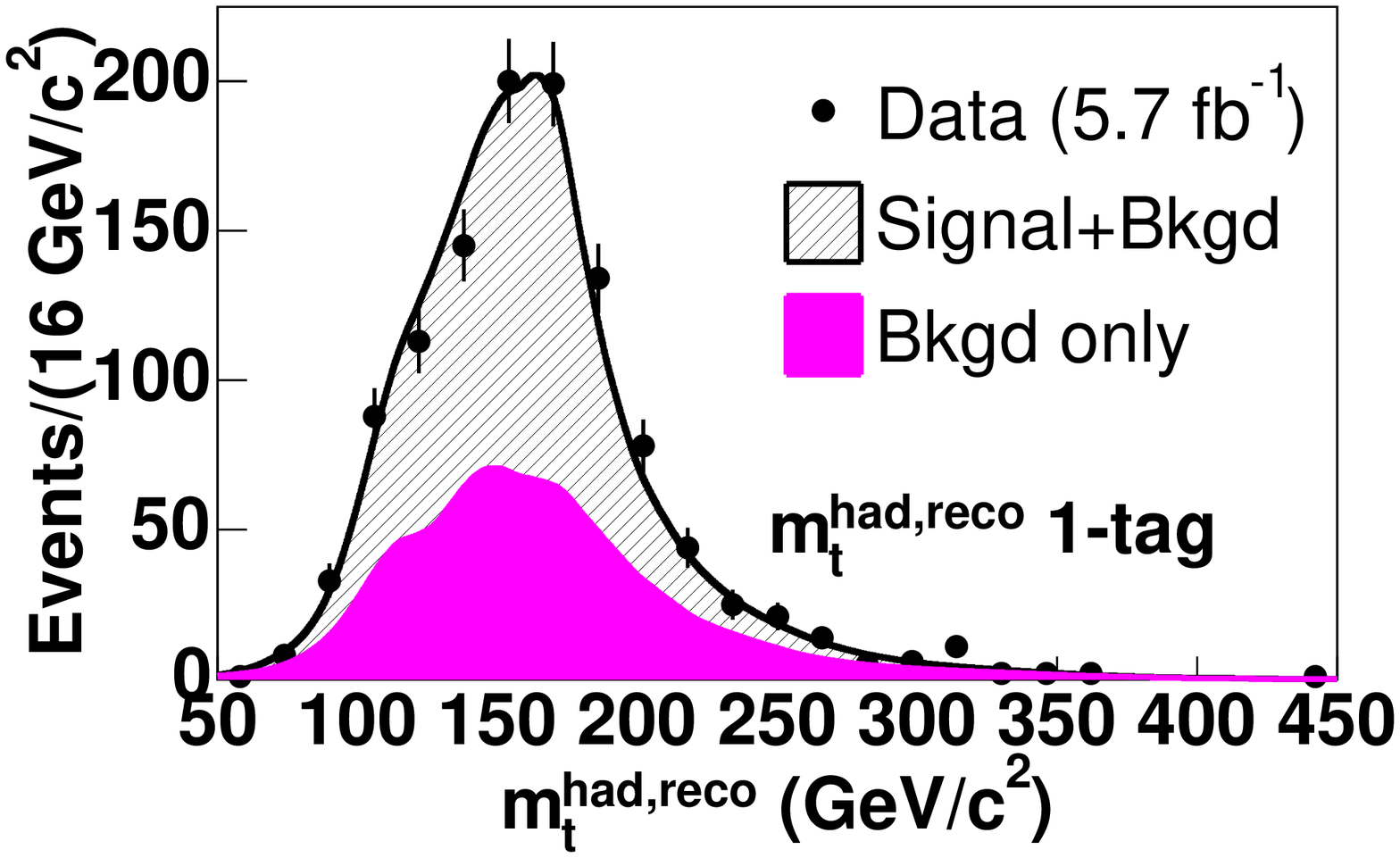} &
\includegraphics[width=0.33\textwidth]{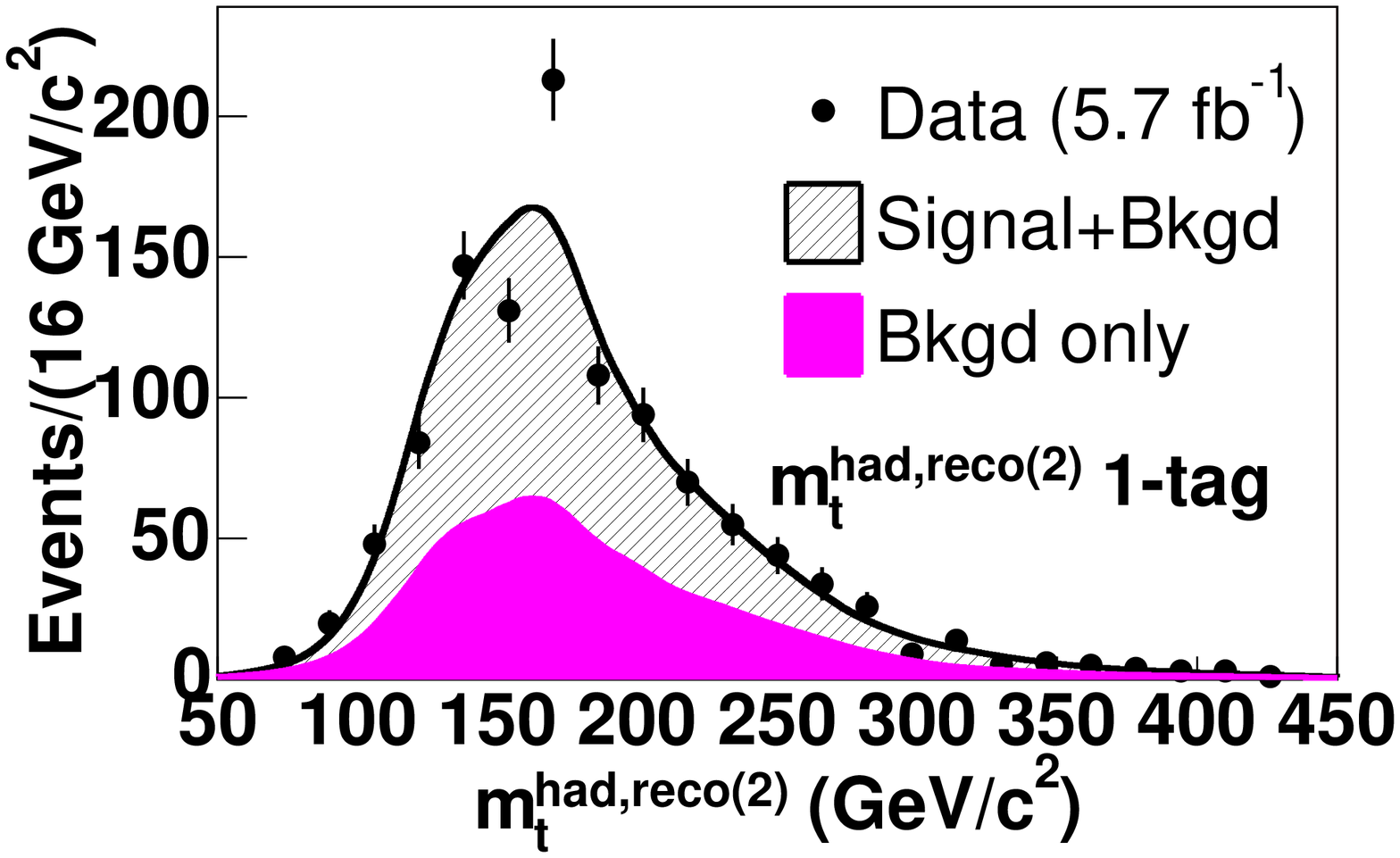}\\
\includegraphics[width=0.33\textwidth]{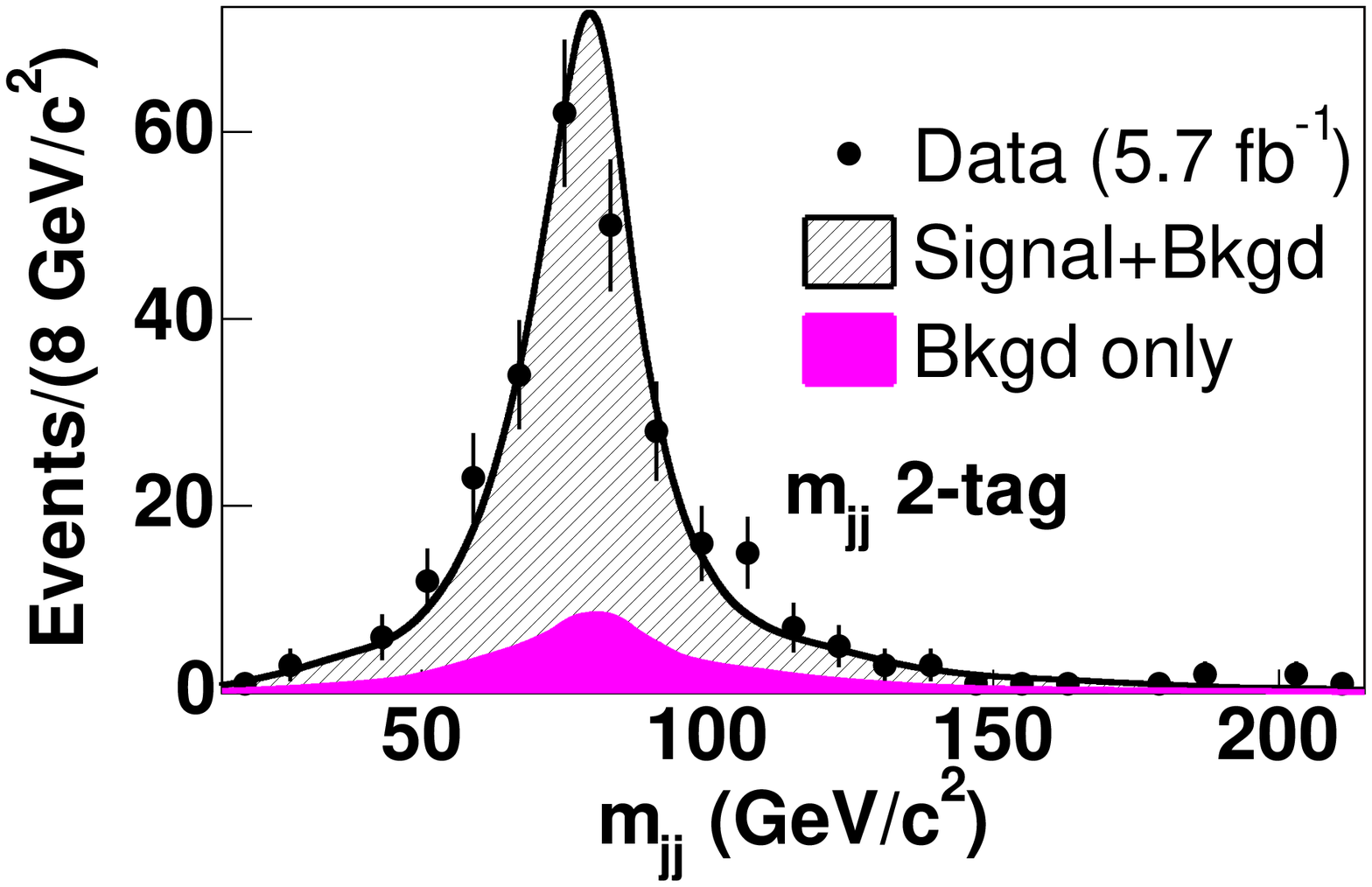} &
\includegraphics[width=0.33\textwidth]{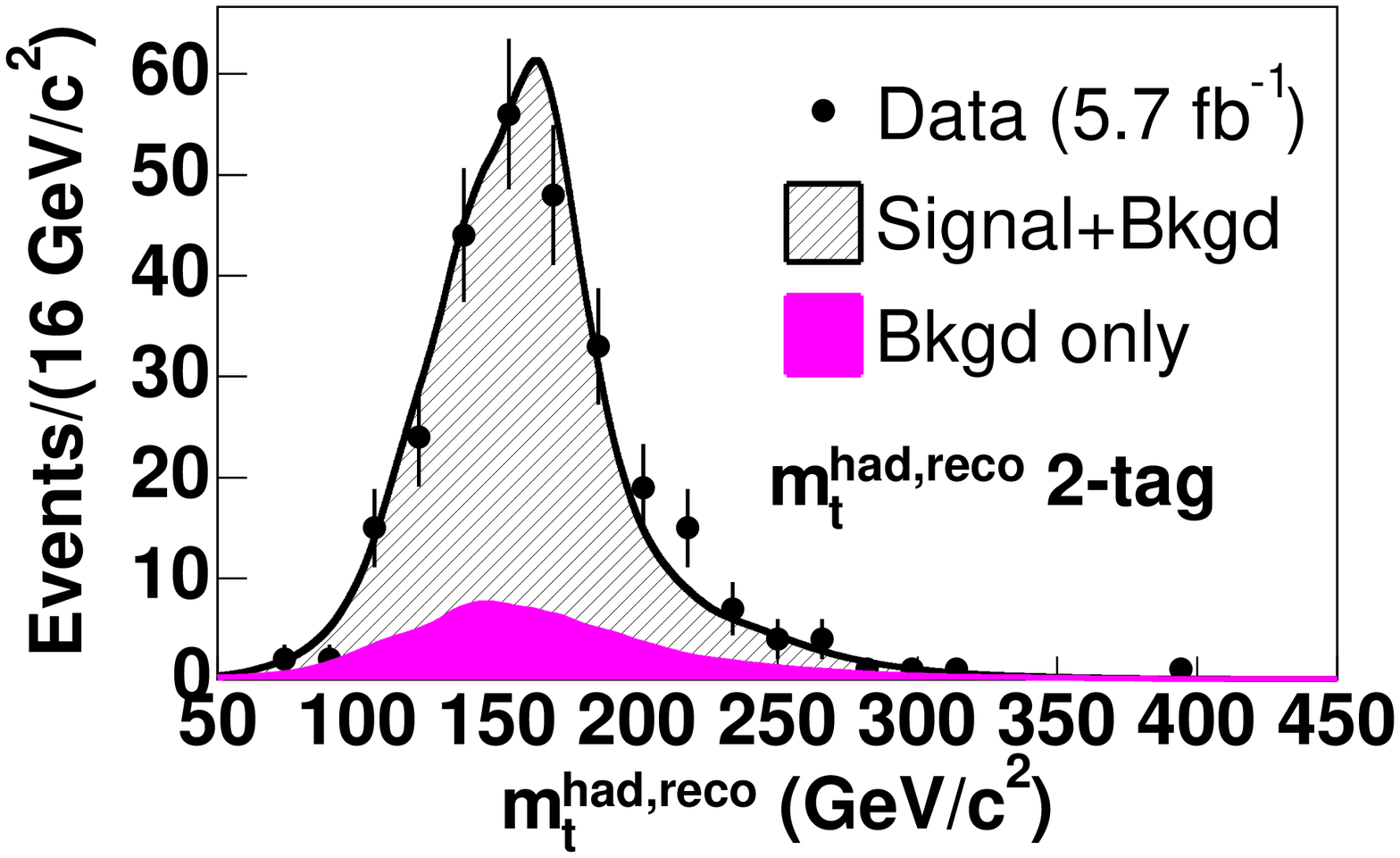} &
\includegraphics[width=0.33\textwidth]{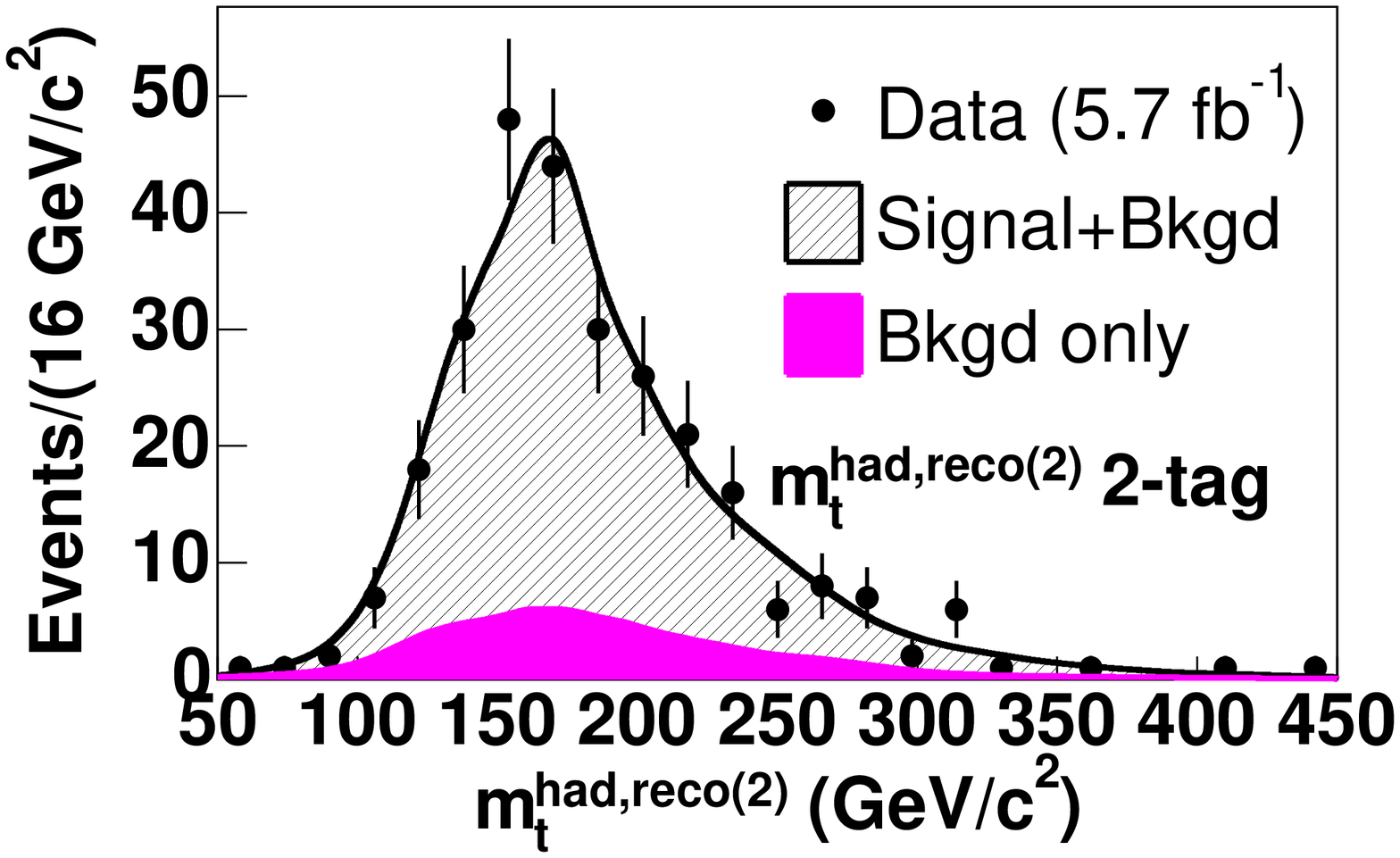} \\
\end{tabular}
\caption[pdff]{Distribution of kinematic variables \mjj, \ma, and \mb from data~(points), overlaid with their corresponding one dimentional p.d.f.s from signal MC sample~(\mtop = 172.5 GeV/c$^2$, hashed area) plus the estimated background~(filled area). Both 1-tag~(top) and 2-tag~(bottom) events are displayed.}
\label{fig:pdf1d}
\end{cfigure1c}

In conclusion, we have performed a measurement of the top-quark mass in events with large \met and jets, corresponding to an integrated luminosity of  5.7 fb$^{-1}$. The data sample has been chosen in such a way as to exclude events used in other CDF top-quark mass measurements. The result, $\mtop = \gevcc{\measErr{172.3}{2.6}}$, is approximately a factor of 6 improvement from the previous measurement in this channel~\cite{earlyMETJ}, and is in agreement with other measurements which contribute to the world average of $\mtop = \gevcc{\measErr{173.2}{0.9}}$~\cite{comb}. 

We thank the Fermilab staff and the technical staffs of the participating institutions for their vital contributions. This work was supported by the U.S. Department of Energy and National Science Foundation; the Italian Istituto Nazionale di Fisica Nucleare; the Ministry of Education, Culture, Sports, Science and Technology of Japan; the Natural Sciences and Engineering Research Council of Canada; the National Science Council of the Republic of China; the Swiss National Science Foundation; the A.P. Sloan Foundation; the Bundesministerium f\"ur Bildung und Forschung, Germany; the Korean World Class University Program, the National Research Foundation of Korea; the Science and Technology Facilities Council and the Royal Society, UK; the Institut National de Physique Nucleaire et Physique des Particules/CNRS; the Russian Foundation for Basic Research; the Ministerio de Ciencia e Innovaci\'{o}n, and Programa Consolider-Ingenio 2010, Spain; the Slovak R\&D Agency; the Academy of Finland; and the Australian Research Council (ARC).


\begin{thebibliography}{99}
\bibitem{DefMET} The transverse momentum $p_T$ and transverse energy $E_T$ of a particle are defined as $|\vec{p}\sin\theta|$ and $E\sin\theta$ repectively, where $\theta$ is the polar angle of the particle momentum with respect to the proton beam direction. The missing transverse energy, an imbalance of energy in the plane transverse to the beam direction, is defined as $\met = |\sum_{towers} E_T{\hat{n}}_T|$ , where ${\hat{n}}_T$ is the unit vector normal to the beam and pointing to a given calorimeter tower and $E_T$ is the transverse energy measured in that tower.
\bibitem{cordSystem} CDF uses a cylindrical coordinate system with the $z$ axis along the proton beam axis. Pseudorapidity is $\eta \equiv -\ln(\tan(\theta/2))$, where $\theta$ is the polar angle relative to the proton beam direction, and $\phi$ is the azimuthal angle while $p_T = |p| \sin\theta$, $E_T = E \sin\theta$.
\end{thebibliography}

\end{document}